\documentclass{PoS}

\newcommand{\e}{\mathrm{e}}

\title{Inflationary universe in a conformally-invariant two scalar-field theory with an $R^2$ term}

\ShortTitle{Inflationary universe in a conformally-invariant two scalar-field theory with an $R^2$ term}

\author{\speaker{Kazuharu Bamba} \\ 
        Faculty of Symbiotic Systems Science, Fukushima University\\
        E-mail: \email{bamba@sss.fukushima-u.ac.jp}}

\abstract{We investigate the inflationary universe in a theory where two scalar fields non-minimally coupling to the scalar curvature and an extra $R^2$ term exist and the conformal invariance is broken. 
In particular, the slow-roll inflation is explored 
for the case that one scalar field is dynamical 
and that two scalar fields are dynamical. 
As a result, we show that the spectral index of the curvature perturbations 
and the tensor-to-scalar ratio of the density perturbations can be compatible with the Planck results. 
It is also demonstrated that the graceful exit from inflation can be realized.}

\FullConference{11th International Workshop Dark Side of the Universe 2015\\
		 14-18 December 2015\\
		 Yukawa Institute for Theoretical Physics, Kyoto University Japan}

\begin{document}

\section{Introduction}

The observations of the Wilkinson Microwave anisotropy probe (WMAP)~\cite{Komatsu:2010fb, Hinshaw:2012aka}, 
the Planck satellite~\cite{Planck:2015xua, Ade:2015lrj}, 
and the BICEP2 experiment~\cite{Ade:2014xna, Ade:2015tva, Array:2015xqh} 
in terms of the anisotropy of the cosmic microwave background (CMB) radiation 
provide the properties of inflation~\cite{Starobinsky:1980te, Sato:1980yn, Guth:1980zm, Linde:1981mu, Albrecht:1982wi}. 
There have been proposed inflation driven by the potential of the single scalar field called the inflation field, e.g., new inflation~\cite{Linde:1981mu, Albrecht:1982wi}, chaotic inflation~\cite{Linde:1983gd}, natural inflation~\cite{Freese:1990rb}, power-law inflation~\cite{Yokoyama:1987an}, and Higgs inflation~\cite{Bezrukov:2007ep}, 
and that of two (multiple) scalar fields or a complex scalar field with two scalar degrees of freedom, for instance, hybrid inflation~\cite{Linde:1993cn} (for reviews on inflationary models, see, for example,~\cite{Lidsey:1995np, Lyth:1998xn, Gorbunov:2011zzc, Martin:2013tda, Linde:2014nna}). 

There also exists 
the Starobinsky inflation ($R^2$ inflation)~\cite{Starobinsky:1980te, Vilenkin:1985md} realized by the $R^2$ term, where $R$ is the scalar curvature. 
It is known that the Starobinsky inflation is consistent with the Planck analysis. This theory can be regarded as a modified gravity theory including $F(R)$ gravity in which the late-time accelerated expansion of the universe can occur (for reviews on the so-called dark energy problem and modified theories of gravitation, see, e.g.,~\cite{Nojiri:2010wj, Nojiri:2006ri, Joyce:2014kja, Book-Capozziello-Faraoni, Capozziello:2011et, Koyama:2015vza, Bamba:2012cp, delaCruzDombriz:2012xy, Bamba:2015uma, Bamba:2013iga, Bamba:2014eea}). 

In this paper, we review the main consequences in Ref.~\cite{Bamba:2015uxa}. 
We explore inflationary cosmology in a theory where there exist 
two scalar fields non-minimally coupling to the scalar curvature 
and an extra $R^2$ term. 
In Ref.~\cite{Bamba:2006mh}, inflation has been considered in 
a theory consisting of two scalar fields which non-minimally couple 
to the scalar curvature. 
The conformally-invariant two-scalar-field theory is studied. 
To add an additional $R^2$ term breaks the conformal invariance. 
Especially, we investigate the slow-roll inflation 
for the case that one scalar field is dynamical 
(i.e., the other scalar field is set to be a constant) 
and that both of two scalar fields are dynamical. 
We examine the spectral index of the curvature perturbations 
and the tensor-to-scalar ratio of the density perturbations. 
The theoretical consequences are compared with 
the observations of the Planck satellite as well as the BICEP2 experiment. 
The spectral index and tensor-to-scalar ratio can be consistent with 
the Planck analysis. 

A unification of inflation in the early universe realized by the $R^2$ term and the late-time accelerated expansion of the universe (namely, 
the dark energy dominated stage) 
is an important motivation of the present theory. 
The $R^2$ term is considered to be a modification of gravity. 
One scalar filed can correspond to dark energy and 
the other scalar can describe dark matter. 
We use units of $k_\mathrm{B} = c = \hbar = 1$ and express the
gravitational constant $8 \pi G_\mathrm{N}$ by
${\kappa}^2 \equiv 8\pi/{M_{\mathrm{Pl}}}^2$ 
with the Planck mass of $M_{\mathrm{Pl}} = G_\mathrm{N}^{-1/2} = 1.2 \times 
10^{19}$\,\,GeV. 

\section{Model}

\subsection{Transformation into the canonical form} 

The action is represented as 
\begin{equation}
S= \int d^4 x \sqrt{-g}\left\{ \frac{\alpha}{2} R^2 + \frac{s}{2}\left[ \frac{(\phi^2-u^2)}{6}R + (\nabla\phi)^2 - (\nabla u)^2 \right] - (\phi^2-u^2)^2J(y) \right\}\,.
\label{eq:2.1}
\end{equation}
Here, $\phi$ and $u$ are two scalar fields, 
$g$ is the determinant of the metric tensor $g_{\mu\nu}$, 
$\alpha (\neq 0)$ is a constant, 
$s=\pm 1$, 
$\nabla$ is the covariant derivative, and $J (y)$ is a function of $y \equiv u/\phi$. 
The action in Eq.~(\ref{eq:2.1}) does not have the conformal invariance, 
but it has the scale invariance. 
Here and in what follows, we have set $2\kappa^2 = 1$. 
When the $R^2$ term does not exist, 
the action in Eq.~(\ref{eq:2.1}) has the conformal invariance. 
If the $R^2$ term is added, 
the conformal invariance is broken. 
We introduce an auxiliary field $\Phi$. 
The action in Eq.~(\ref{eq:2.1}) can be expressed as 
\begin{equation}
S= \int d^4 x \sqrt{-g}\left\{ \left[ \Phi + \frac{s}{12}(\phi^2-u^2) \right]R -\frac{\Phi^2}{2\alpha} + \frac{s}{2} \left[  (\nabla\phi)^2 - (\nabla u)^2 \right] - (\phi^2-u^2)^2J(y) \right\}\,.
\label{eq:2.2}
\end{equation}
By taking the gauge 
$\Phi + \left(s/12\right)\left(\phi^2-u^2\right) = 1$, 
the action in Eq.~(\ref{eq:2.2}) is transformed into  
the canonical form. 
Namely, 
the non-minimal coupling of the scalar fields to 
the scalar curvature is removed. 
Accordingly, the Einstein-Hilbert term appears and 
the scale invariance is broken. 
As a result, the action in Eq.~(\ref{eq:2.2}) is described as 
\begin{eqnarray}
&&
S = \int d^4 x \sqrt{-g}\left\{ R  + \frac{s}{2} \left[ (\nabla\phi)^2 - (\nabla u)^2 \right] - V(\phi, u, J) \right\}\,. 
\label{eq:2.3}\\ 
&&
V(\phi, u, J) \equiv \frac{1}{2\alpha}\left[ 1 - \frac{s}{12}(\phi^2-u^2) \right]^2+(\phi^2-u^2)^2 J(y) \,, 
\label{eq:2.4}
\end{eqnarray}
where $V(\phi, u, J)$ is a potential for $u$ and $\phi$. 
It follows from the action in Eq.~(\ref{eq:2.3}) that 
the gravitational field equation is derived as 
\begin{equation}
R_{\mu\nu}-\frac{1}{2}Rg_{\mu\nu}+\frac{s}{2}\left(\nabla_{\mu}\phi\nabla_{\nu}\phi-\nabla_{\mu} u \nabla_{\nu} u\right) + \frac{1}{2}g_{\mu\nu} \left\{ V - \frac{s}{2}\left[(\nabla\phi)^2-(\nabla u)^2\right]
\right\} =0 \,. 
\label{eq:2.5}
\end{equation}
Moreover, the equations of motion for $\phi$ and $u$ are given by 
\begin{eqnarray}
&&
s\Box\phi + V_{\phi} = 0\,, 
\label{eq:2.6} \\  
&&
s\Box u - V_{u} = 0\,, 
\label{eq:2.7}
\end{eqnarray}
where $V_{\phi} \equiv \partial V/\partial \phi$ and 
$V_{u} \equiv \partial V/\partial u$. 

We suppose 
the flat Friedmann-Lema\^{i}tre-Robertson-Walker (FLRW) metric 
$ds^2 = -dt^2 + a^2(t) \sum_{i=1,2,3}\left(dx^i\right)^2$ 
with $a(t)$ the scale factor. 
The Hubble parameter is defined as 
$H \equiv \dot{a}/a$. Here,  
the dot denotes the time derivative.  
In the flat FLRW background, 
the gravitational field equations are given by 
\begin{eqnarray}
&&
3H^2 + \frac{s}{4}(\dot\phi^2-\dot u^2)-\frac{1}{2}V=0 \,,
\label{eq:2.8} \\ 
&&
2\dot H + 3H^2 - \frac{s}{4}(\dot\phi^2-\dot u^2)-\frac{1}{2}V=0 \,. 
\label{eq:2.9}
\end{eqnarray}
Furthermore, the equations of motion for $\phi$ and $u$ read 
\begin{eqnarray} 
&&
s(\ddot\phi+3H\dot\phi) - V_{\phi} = 0\,,
\label{eq:2.10} \\ 
&&
s(\ddot{u}+3H\dot{u}) + V_{u} = 0\,.
\label{eq:2.11}
\end{eqnarray}
%

\subsection{Case that the single scalar field is dynamical}

We study the case that $\phi$ is the inflaton field. 
If $\left|\ddot{\phi}\right| \ll \left| 3H \dot{\phi} \right|$ 
in the equation of motion for $\phi$ (\ref{eq:2.10}) 
and the kinetic energy $\left(1/2\right) \dot{\phi}^2$ of $\phi$ is 
much smaller than the potential $V$, 
the slow-roll inflation happens. 
In addition, it is assumed that the mass of $u$ is much smaller than the Hubble parameter during inflation, and therefore the value of $u$ does not change at the inflationary stage. 
Hence, we take $u=u_0 (= \mathrm{constant})$ during inflation and $J=J(y^2)$. 
As the simplest case, we select $u_0=0$ and $J=1$.  
In this case, $u$ is completely decoupled to equations 
at the inflationary stage. 
Accordingly, 
the effective potential of $\phi$ becomes 
\begin{equation}
V_\mathrm{eff}(\phi) = \frac{1}{2\alpha}
\left(1 - \frac{s}{12}\phi^2 \right)^2+C\phi^4 \,, 
\label{eq:2.12}
\end{equation}
where $C$ is a constant. 
This representation is available for any function of $J=J(y^2)$. 
The number of $e$-folds at the inflationary stage is given by 
\begin{equation}
N_e(\phi)=\int^{\phi_\mathrm{f}}_{\phi}H(\hat{\phi})\frac{d\hat{\phi}}{\dot{\hat{\phi}}}
=-\frac{s}{2}\int^{\phi}_{\phi_\mathrm{f}}\frac{V(u, \hat{\phi}, J)}{V_\phi (u, \hat{\phi}, J)}d\hat{\phi} \,,
\label{eq:2.13}
\end{equation}
with $\phi_\mathrm{f}$ the value of $\phi$ at the end time $t_\mathrm{f}$ of inflation. 

In the kinematic approach, the slow-roll parameters are defined by 
$\epsilon \equiv -\dot{H}/H^2$, 
$\eta \equiv \epsilon - \ddot{H}/\left(2H \dot{H}\right)$ (for reviews, see, e.g.,~\cite{Lidsey:1995np, Lyth:1998xn}). 
In the case of the slow-roll inflation, 
the spectral index $n_\mathrm{s}$ of the curvature perturbations and 
the tensor-to-scalar ratio $r$ of the density perturbations 
are written as~\cite{Mukhanov:1981xt, L-L}  
$n_\mathrm{s} - 1 = -6\epsilon + 2\eta$ and 
$r = 16 \epsilon$, respectively. 

The Planck analysis has shown 
$n_{\mathrm{s}} = 0.968 \pm 0.006\, (68\%\,\mathrm{CL})$~\cite{Planck:2015xua, Ade:2015lrj} and $r < 0.11\, (95\%\,\mathrm{CL})$~\cite{Ade:2015lrj}, 
which are compatible with the WMAP results~\cite{Komatsu:2010fb, Hinshaw:2012aka}. The BICEP2 experiment and Keck Array have indicated 
$r<0.09\, (95\%\,\mathrm{CL})$~\cite{Array:2015xqh}. 

In this model, 
for $N_e = 55$, which are large enough to 
explain the horizon and flatness problems, 
$n_\mathrm{s} = 0.9603$, and $s=-1$, 
we have $\phi \approx 34.7$, $x \equiv \alpha C \approx 2.8 \times 10^9$, and 
consequently acquire $r \approx 0.212$.

\section{Case that two scalar fields are dynamical}

Next, in the theory whose action is given by Eq.~(\ref{eq:2.1}), 
we study inflation for the case that both of two scalar fields $\phi$ and $u$ 
are dynamical and they are non-interacting scalar fields~\cite{Starobinsky:1986fxa}. 
The expressions of the slow-roll parameters 
and those of the spectral index $n_\mathrm{s}$ of the curvature perturbations 
and the tensor-to-scalar ratio $r$ of the density perturbations 
for the case of the single dynamical scalar field shown in Sec.~2 
can be used also for the case of the two dynamical scalar fields~\cite{Achucarro:2015bra}. 

In the flat FLRW space-time, the system of equations is 
given by Eqs.~(\ref{eq:2.8})--(\ref{eq:2.11}). 
By assuming the slow-roll conditions 
$\ddot u \ll \dot u H$, $\ddot\phi\ll\dot\phi H$, 
and $\dot\phi^2-\dot u^2\ll H^2$, 
we find the Friedmann equation $H^2 = \left(1/6\right) V$ 
and the equations of motion for $\phi$ and 
$u$, $3sH \dot{\phi} = V_{\phi}$ and 
$3sH \dot{u} = - V_u$. 
The number of $e$-folds is described by 
\begin{equation}
N_{e}\equiv\int^{a_\mathrm{f}}_{a_\mathrm{i}}d\ln a= \int^{t_\mathrm{f}}_{t_\mathrm{i}}H dt 
= \frac{s}{2}\int_{\phi,u}^{\phi_\mathrm{f}, u_\mathrm{f}}\frac{V \left(V_u du+V_{\phi}d\phi\right)}{V_{\phi}^2-V_u^2}\,. 
\label{eq:3.1}
\end{equation}
Here, $a_\mathrm{i}$ and $a_\mathrm{f}$ are the values of $a(t)$ at the initial time $t_\mathrm{i}$ and the end time $t_\mathrm{f}$ of inflation, 
respectively. Furthermore, $\phi_\mathrm{f}$ and $u_\mathrm{f}$ are the values of $\phi$ and $u$ at the end time $t_\mathrm{f}$ of inflation, respectively. 

The slow-roll parameters  
$\epsilon$ and $\eta$ are represented as 
\begin{equation}
\epsilon = -\frac{\dot V}{2HV}= \frac{V_u^2 - V_{\phi}^2}{sV^2} \,, 
\quad 
\eta = -\frac{1}{4HV\dot V} \left(\dot V^2 + 2\ddot VV \right) 
=-\frac{2\left(V_{\phi}^2V_{\phi\phi}+V_u^2V_{uu}\right)}{sV\left( V_{\phi}^2-V_u^2 \right)} \,,
\label{eq:3.2}
\end{equation}
with $V_{\phi\phi} \equiv \partial^2 V/\partial \phi^2$ and 
$V_{uu} \equiv \partial^2 V/\partial u^2$. 

We take $J=C$ and $s=-1$. 
If $N_e=50$ and $x=10^{30}$, we obtain 
$\phi^2-u^2=2593$, and eventually we find $r\simeq 0.1$ and 
$n_\mathrm{s}=3.7 \times 10^{-9} \mathcal{G}^4+0.9815$, 
where $\mathcal{G}^4 \equiv u^2\phi^2$. 
In addition, when $N_e=60$ and $x=10^{25}$, we have 
$\phi^2-u^2=2317$, and finally we get $r\simeq 0.11$ and 
$n_\mathrm{s} = 5.0 \times 10^{-9} \mathcal{G}^4 + 0.9793$. 
Thus, $n_\mathrm{s}$ and $r$ cannot be consistent with 
the Planck results.

\section{Inflationary cosmology in the Einstein frame}

The action in Eq.~(\ref{eq:2.1}) is described in the Jordan frame. 
We introduce an auxiliary field $\Phi$ and 
execute the conformal transformation of the metric from the Jordan frame to the Einstein frame~\cite{Maeda:1988ab, F-M} 
as $g_{\mu\nu}=\Lambda \bar{g}_{\mu\nu}$ with 
$\Lambda \equiv \e^{\lambda}$, where $\lambda$ is a scalar field. 
Here, the bar means the quantities in the Einstein frame. 
In this case, we have $\sqrt{-g}=\Lambda^2\sqrt{-\bar{g}}$. 
In the action in Eq.~(\ref{eq:2.2}), we set 
$\Lambda\left[\Phi+\left(s/12\right)\left(\phi^2-u^2\right)\right]=1$. 
Accordingly, we acquire
\begin{eqnarray}
&&
S = \int d^4x\sqrt{-\bar{g}}\left\{ \bar{R} -\frac{3}{2}\left(\bar{\nabla}\lambda\right)^2 + \frac{s}{2}\e^{\lambda} \left[\left(\bar{\nabla}\phi\right)^2 - \left(\bar{\nabla} u\right)^2 \right] -  V(\lambda,\phi,u,J) \right\}\,.
\label{eq:4.1} \\
&&
V(\lambda,\phi,u,J) = \frac{1}{2\alpha}\left[1-\frac{s}{12}\e^{\lambda}(\phi^2-u^2)\right]^2 + \e^{2\lambda}(\phi^2-u^2)^2J(y)\,, 
\label{eq:4.2}
\end{eqnarray}
where $V(\lambda,\phi,u,J)$ is the potential for the scalar fields $\lambda$, $\phi$, and $u$. 
In the following, we will not describe the bar over the quantities and operators in the Einstein frame for simplicity. 
{}From the action in Eq.~(\ref{eq:4.1}), the gravitational field equation 
is found as 
\begin{eqnarray}
&&
R_{\mu\nu}-\frac{1}{2}Rg_{\mu\nu}+\frac{s}{2}\e^{\lambda}\left(\nabla_{\mu}\phi\nabla_{\nu}\phi-\nabla_{\mu} u \nabla_{\nu} u\right) -\frac{3}{2}\nabla_{\mu}\lambda\nabla_{\nu}\lambda 
\nonumber \\
&&
{}+\frac{1}{2}g_{\mu\nu} \left\{ V - \frac{s}{2}\e^{\lambda}\left[ \left(\nabla\phi\right)^2-\left(\nabla u\right)^2 \right] +\frac{3}{2}\left(\nabla\lambda\right)^2 \right\} =0\,. 
\label{eq:4.3}
\end{eqnarray}
Furthermore, the equations of motion for $\lambda$, $\phi$, and $u$ become
\begin{eqnarray}
&&
3\Box\lambda +\frac{s}{2}\e^{\lambda}\left[\left(\nabla\phi\right)^2-\left(\nabla u\right)^2\right] - V_{\lambda}=0\,,
\label{eq:4.4} \\
&&
s\e^{\lambda}\Box\phi + s\e^{\lambda}\nabla^{\mu}\lambda\nabla_{\mu}\phi + V_{\phi}=0\,,
\label{eq:4.5} \\ 
&&
s \e^{\lambda}\Box u 
+ s\e^{\lambda}\nabla^{\mu}\lambda\nabla_{\mu}u - V_{u}=0\,. 
\label{eq:4.6}
\end{eqnarray}

In the flat FLRW space-time, from Eq.~(\ref{eq:4.3}), 
the gravitational field equations read 
\begin{eqnarray}
&&
3H^2 + \frac{s}{4}\e^{\lambda}\left(\dot\phi^2-\dot u^2\right) -\frac{3}{4}\dot\lambda^2-\frac{1}{2}V=0\,,
\label{eq:4.7} \\
&&
2\dot H + 3H^2 - \frac{s}{4}\e^{\lambda}\left(\dot\phi^2-\dot u^2\right) +\frac{3}{4}\dot\lambda^2-\frac{1}{2}V=0\,. 
\label{eq:4.8}
\end{eqnarray}
Moreover, from Eqs.~(\ref{eq:4.4})--(\ref{eq:4.6}), 
the equations of motion for $\lambda$, $\phi$, and $u$ are given by 
\begin{eqnarray}
&&
3\ddot\lambda + 9H\dot\lambda +\frac{s}{2}\e^{\lambda}\left(\dot\phi^2-\dot u^2\right) + V_{\lambda}=0\,,
\label{eq:4.9} \\
&&
s\e^{\lambda}\ddot\phi + 3s\e^{\lambda}H\dot\phi + s\e^{\lambda}\dot\lambda\dot\phi - V_{\phi}=0\,,
\label{eq:4.10} \\
&&
s\e^{\lambda}\ddot u + 3s\e^{\lambda}H\dot u + s\e^{\lambda}\dot\lambda\dot u + V_{u}=0\,,
\label{eq:4.11}
\end{eqnarray}
where $V_{\lambda} \equiv \partial V/\partial \lambda$

We explore the case that the scalar field 
$\lambda$ is the inflaton field and other two scalar fields 
$\phi$ and $u$ are taken to be non-zero constants as 
$\phi=\phi_0 (= \mathrm{constant} \neq 0)$ and $u=u_0 (= \mathrm{constant} \neq 0)$, where $u_0 \neq \pm \phi_0$. 
We set $J(y)=C\left(y-y_0\right)^q$ 
with $q (\geq 2)$ a constant, where we have   
$J(y_0)=0$ and $dJ(y_0)/dy=0$. 
The effective potential is given by 
$V_\mathrm{eff} (\lambda) = \left[1/\left(2\alpha\right)\right] 
\left[1-\left(\zeta/12\right) \e^{\lambda}\right]^2$ 
with $\zeta \equiv s(\phi_0^2-u_0^2)$. 
We also get $N_{e}(\phi)=\left(3/4\right)\lambda + 
\left[9/\left(\zeta \e^{\lambda}\right)\right]$. 
Hence, we obtain 
\begin{equation}
n_\mathrm{s} = \frac{432 - 5 \zeta^2 \e^{2\lambda}-168\zeta \e^{\lambda}}{3\left(\zeta \e^{\lambda}-12\right)^2}\,, 
\quad 
r = \frac{64 \zeta^2 \e^{2\lambda}}{3\left(\zeta \e^{\lambda}-12\right)^2} \,.
\label{eq:4.11}  
\end{equation}
For $N_e=60$ and $\zeta = 0.10$, we acquire $n_\mathrm{s}=0.9652$ and 
$r=4.0 \times 10^{-3}$. These are consistent with the Planck results.

\section{Instability of the de Sitter solution}

In the case that one scalar field is dynamical in the Einstein frame 
considered in Sec.~4, 
we examine the instability of the de Sitter solution 
to describe inflation and study the graceful exit from inflation.  
 
We take the perturbations of the Hubble parameter during inflation as 
$H = H_\mathrm{inf} \left( 1 + \delta(t) \right)$ 
with $\left| \delta(t) \right| \ll 1$. 
Here, $H_\mathrm{inf} (>0)$ (= constant) 
is the Hubble parameter at the inflationary stage, 
and $\delta(t)$ shows the perturbations from the de Sitter solution. 
We consider the case that $\lambda$ is only dynamical 
and $\phi$ and $u$ are constants. 
By making the time derivative of Eq.~(\ref{eq:4.8}) 
with $V = V_\mathrm{eff}$, we have 
\begin{equation}
2\ddot{H} +6H\dot{H} +\frac{3}{2} \dot{\lambda} \ddot{\lambda} + 
\frac{\zeta}{24\alpha} \left(1-\frac{\zeta}{12}
\e^{\lambda} \right)\e^{\lambda} \dot{\lambda} = 0 \,.
\label{eq:5.1} 
\end{equation}
It follows from Eq.~(\ref{eq:4.9}) with $V = V_\mathrm{eff}$ 
that the equation of motion for $\lambda$ is written as 
\begin{equation} 
3\ddot{\lambda} + 9H\dot{\lambda} - \frac{\zeta}{12\alpha} \left( 
1-\frac{\zeta}{12} \e^{\lambda} \right) \e^{\lambda} =0\,. 
\label{eq:5.2} 
\end{equation}
Here, the limit $t \to 0$ is taken because we study inflation in the early universe. For this limit, an approximate solution $\lambda \approx \ln \left( H t \right)$ for Eq.~(\ref{eq:5.2}) and the quasi de Sitter solution 
$H \approx 1/\left(3t \right)$ could be found. 
Therefore, in this limit, 
the third term $\left(3/2\right) \dot{\lambda} \ddot{\lambda}$ 
in the left-hand side of Eq.~(\ref{eq:5.1}) 
is much larger than the fourth term proportional only to 
$\dot{\lambda}$. 
Thus, the fourth term would be negligible. 

We express $\delta(t)$ as $\delta(t) = \e^{\beta t}$, 
where $\beta$ is a constant. If $\beta >0$, 
the de Sitter solution representing inflation is unstable, 
and eventually the graceful exit from inflation 
can be realized and the reheating stage can follow it. 
Namely, when $\beta$ is positive, 
the value of $\delta(t)$ becomes large in time. 
We substitute the representation of perturbations of $H = H_\mathrm{inf} \left( 1 + \delta(t) \right)$ with $\delta(t) = \e^{\beta t}$ 
into Eq.~(\ref{eq:5.1}) and use the approximate solutions $\lambda \approx 
\ln \left( H t \right)$ and $H \approx 1/\left(3t \right)$. 
As a result, we acquire 
$2H_\mathrm{inf} \beta^2 + 6H_\mathrm{inf}^2 \beta + \left(81/2\right) 
H_\mathrm{inf}^3 = 0$. 
The solutions of this equation are given by 
$\beta_\pm = \left[3\left(-1\pm\sqrt{10}\right)H_\mathrm{inf}\right]/2$. 
Thus, the positive solution of $\beta = \beta_+ > 0$ is obtained. 
Consequently, the universe can gracefully exit from inflation.

\section{Summary}

We have considered inflation 
in a theory consisting of 
two scalar fields which non-minimally couple to the scalar curvature and 
an extra $R^2$ term. 
We have studied the slow-roll inflation 
in the case that the single scalar field is dynamical and 
that both of two scalar fields are dynamical. 
We have investigated the spectral index $n_\mathrm{s}$ of the curvature perturbations and the tensor-to-scalar ratio $r$ of the density perturbations 
and compared the theoretical results with the observational ones of the Planck satellite and the BICEP2 experiment. 
In the Jordan frame, for both of the cases that the single scalar field 
is dynamical and that two scalar fields are dynamical, 
when the number of $e$-folds at the inflationary stage is given by 
$50 \leq N_e \leq 60$, 
we obtain $n_\mathrm{s} \approx 0.96$ and $r = \mathcal{O} (0.1)$. 
While, in the Einstein frame, 
for the case that one scalar field is dynamical, 
we find $n_\mathrm{s} \approx 0.96$ and $r < 0.11$, 
which are compatible with the Planck analysis.  
Thus, in this theory, 
inflation with the spectral index and the tensor-to-scalar ratio consistent with the Planck results can be realized. 
In addition, it has been demonstrated that 
the de Sitter solution is unstable, and hence the graceful exit from 
inflation can be realized.

\section*{Acknowledgments}

The author would like to thank Professor Sergei D. Odintsov and 
Dr. Petr V. Tretyakov for our collaboration in our work~\cite{Bamba:2015uxa} 
very much. 
This work was partially supported by the JSPS Grant-in-Aid for 
Young Scientists (B) \# 25800136 and 
the research-funds presented by Fukushima University.

\end{document}